%% file: main_v2.tex
\documentclass[
    twocolumn,
	prd,
	amssymb,
	preprintnumbers,superscriptaddress,
	nofootinbib]{revtex4-1}

\pdfoutput=1
\usepackage{paralist}
\usepackage{graphicx}
\usepackage{enumitem}
\usepackage{latexsym}
\usepackage{amsfonts}
\usepackage{amssymb}
\usepackage{xcolor}
\usepackage[export]{adjustbox}
\usepackage{amsmath}
\usepackage[thinlines]{easytable}
\usepackage{slashed}
\usepackage{dcolumn}
\usepackage{verbatim}
\usepackage{float}
\usepackage{multirow}
\usepackage{xspace}
\usepackage[normalem]{ulem}
\usepackage[
pdfauthor={Djuna Croon}]{hyperref}
\usepackage{tabularx}
\usepackage{lettrine}
\usepackage{setspace}

\input Zallman.fd

\LettrineTextFont{\itshape}

\input{universalnewcommands.tex}

\newcommand{\vtwo}[1]{\textcolor[rgb]{0.,0,0.}{#1}}

\allowdisplaybreaks

\begin{document}

\title{Caustic crossings in giant arcs with extended dark matter objects}

\author{Djuna Croon} \email{djuna.l.croon@durham.ac.uk}
\affiliation{Institute for Particle Physics Phenomenology, Department of Physics, Durham University, Durham DH1 3LE, U.K.}
\author{Benedict Crossey} \email{benedict.g.crossey@durham.ac.uk}
\affiliation{Institute for Particle Physics Phenomenology, Department of Physics, Durham University, Durham DH1 3LE, U.K.}
\author{Jose Maria Diego} 
\affiliation{Instituto de Fisica de Cantabria (CSIC-UC), Avda. Los Castros s/n, 39005 Santander, Spain}
\author{Bradley J. Kavanagh} \email{kavanagh@ifca.unican.es}
\affiliation{Instituto de Fisica de Cantabria (CSIC-UC), Avda. Los Castros s/n, 39005 Santander, Spain}
\author{Jose Maria Palencia} 
\affiliation{Instituto de Fisica de Cantabria (CSIC-UC), Avda. Los Castros s/n, 39005 Santander, Spain}

\date{\today}

\begin{abstract}
Caustic-crossing stars observed in giant arcs behind galaxy clusters provide a powerful probe of dark matter substructure. While previous work has focused on point-like lenses such as primordial black holes, we extend this framework to \emph{extended dark objects} (EDOs), including ultracompact minihalos formed from the collapse of primordial overdensities. We develop an analytic model of microlensing by EDOs embedded in a macrolensing cluster potential and derive the resulting caustics and light curves. Depending on the EDO size relative to the effective Einstein radius, we show that they may generate additional narrow caustics, leading to novel features in the light curve. Applying our framework to the MACS J1149 LS1 ``Icarus'' event, we constrain EDOs with radii up to $10^7 R_\odot$. Our results demonstrate that caustic-crossing events complement galactic microlensing searches, as they can probe EDOs with larger physical size. We discuss the implications for current and future observations, which promise to deliver a statistical sample of caustic transients and correspondingly sharper constraints on dark objects.
\end{abstract}

\preprint{IPPP/25/82}

\maketitle

\section{Introduction}

The discovery of the highly magnified star MACS~J1149 LS1 (``Icarus'') at $z=1.49$~\cite{Kelly:2014mwa,Kelly:2017fps,Diego:2017drh} marked the first time that an individual star at cosmological distance was observed through gravitational lensing. 
Such events become possible when the image of a background star lies extremely close to the critical curve of a galaxy cluster. 
The critical curve is the region of the lens plane where the magnification is maximal (formally diverging for a point-like source) and can be mapped onto the corresponding caustic curve in the source plane. 
The smooth cluster potential alone produces a fold caustic, but the actual detection of a caustic-crossing star requires the presence of a microlens -- such as an intracluster (ICL) star or compact object -- that perturbs the macro caustic and generates the microcaustics responsible for observable variability. The resulting light curve encodes information about the small-scale matter distribution near the critical curve, and in particular about the abundance and nature of compact lenses. This opens a new avenue to probe dark objects~\cite{Diego:2017drh,Oguri:2017ock,Palencia:2023kne,Muller:2024pwn}. Since the Icarus observation, several additional caustic-crossing stars have been identified in giant arcs (e.g.~Refs.~\cite{2022Natur.603..815W,welch2022jwst,Fudamoto:2024idh}), establishing these events as a recurring phenomenon and strengthening their role as a probe of dark matter (DM).

As we show in this work, caustic crossing events in giant arcs complement traditional galactic microlensing surveys as they extend sensitivity to \emph{extended} dark objects (EDOs) with sizes that are otherwise inaccessible, \vtwo{although still on a scale much smaller than the cluster}.
So far, most work has focused on point-like lenses such as primordial black holes (PBHs)~\cite{Carr:2020gox,Green:2020jor,Green:2024bam}, for which the expected signatures are well understood. However, a wide range of motivated DM objects -- such as ultracompact minihalos~\cite{Berezinsky:2003vn,Ricotti:2009bs,Scott:2009tu}, boson stars~\cite{JETZER1992163,Dabrowski:1998ac}, axion miniclusters~\cite{HOGAN1988228,Kolb:1993hw,Kolb:1993zz,Fairbairn:2017dmf,Fairbairn:2017sil}, and other EDOs -- have finite radii that significantly alter the microlensing behaviour
\vtwo{, while remaining a small perturbation to the underlying cluster lens, as in the point-like case.}

The Einstein radius sets the characteristic scale for gravitational lensing: 
lenses much smaller than this behave as point masses, objects of comparable size can generate distinctive new lensing features, while those much larger are too diffuse to produce appreciable magnification.
As a result, typical microlensing searches quickly lose sensitivity as the size of the EDO is increased~\cite{Croon:2020wpr,Croon:2020ouk,Fujikura:2021omw,CrispimRomao:2024nbr}.
However, for caustic crossings in giant arcs, the Einstein radius is effectively enlarged both by the strong tangential magnification of the cluster macrolens and by the fact that the lensing distances are cosmological rather than kiloparsec-scale. This makes caustic-crossing stars in giant arcs sensitive to much larger EDOs than typical microlensing searches, probing DM structure on sub-solar to intermediate mass scales \cite{Bringmann:2025cht}.

In this work, we develop an analytic framework to describe microlensing by EDOs embedded in a cluster potential, generalising previous treatments of point lenses. We compute the resulting caustic structures and light curves, quantify the efficiency of extended lenses, and derive constraints on their abundance from the Icarus event. 
We finish by describing the potential of future observations to improve these constraints. 
Our results demonstrate that cluster caustic crossings probe a distinct region of EDO parameter space, providing a powerful complement to existing microlensing searches.

\section{Analytic model}\label{sec:model}
We will first discuss some general properties of extended lenses in a macrolens background, by generalising the treatment of a point source with a point microlens in \cite{Chang:1979zz,Paczynski:1986aa,Kayser:1986aa,Witt:1990aa,miralda1991magnification,Schechter:2002dm,Oguri:2017ock} often referred to as the Chang--Refsdal configuration. For the extended microlens, we follow a similar parametrisation as was used for galactic microlensing in \cite{Croon:2020wpr,Croon:2020ouk}. 
We define the normalised lens-plane-projected mass profile $m(\tau) \equiv M(\tau \theta_E)/M$ where
\begin{equation}
\begin{split}
    M(\theta)=& 2\pi D_{\rm L}^2\int_0^\theta d\theta^\prime \theta^\prime \Sigma(\theta^\prime)~,\\
\Sigma(\theta)=&\int_{-\infty}^\infty dz\,\rho\left(\sqrt{D_{\rm L}^2\theta^2+z^2}\right),
\end{split}
\label{eq:Mthetasigmatheta} 
\end{equation}
$\theta_E = \sqrt{4GM D_{\rm LS}/ D_{\rm L}D_{\rm S} c^2} $ is the point-like Einstein angle, $M\equiv M(\infty)$ is the total microlens mass, and $\Sigma(\theta)$ is the lens-plane-projected surface mass density. We normalise angles by the Einstein angle as this will simplify our equations, $\tau \equiv \theta/\theta_E$.
For the macrolens, we assume constant convergence ($\bar\kappa$) and shear ($\bar\gamma$), such that
\begin{align}
\mu_t^{-1} &= 1 - \bar{\kappa} - \bar{\gamma},  \\
\mu_r^{-1} &= 1 - \bar{\kappa} + \bar{\gamma}, 
\end{align} 
are the tangential (perpendicular to the radial line from the lens centre) and radial macro-magnifications respectively. Then, the total macro-magnification in the absence of a microlens is given by $\bar\mu = \mu_r \mu_t$.\footnote{In practice, the macro model magnifications $\mu_t$ and $\mu_r$
depend on the image position with respect to the critical curve of the macro model.
In particular, one expects $\mu_t \propto \theta_h^{-1}$ where $\theta_h$ is the distance between the image and the critical curve. $\mu_r$ is approximately constant near the critical curve. }
Under this assumption, the microlensing equations for an extended lens are given by
\begin{equation}
    \begin{split}
        u_1 &= \frac{\tau_1}{\mu_r} - \frac{\tau_1}{\tau^2} m(\tau) \\
        u_2 &= \frac{\tau_2}{\mu_t} - \frac{\tau_2}{\tau^2} m(\tau)
    \end{split}
    \label{eq:lensingeqns}
\end{equation}
where $u_i \equiv \beta_i/\theta_E$ is the normalised impact angle and $\tau  = \sqrt{\tau_1^2 + \tau_2^2}$. Thus, in these normalised coordinates, $(u_1,u_2)$ is the position of the source and $(\tau_1,\tau_2)$ that of the image(s), with the centre of the lens taken as the origin of the coordinates.

From the lensing equations one can find the inverse magnification matrix given by
\begin{equation}\begin{split}
    \frac{\partial u_i}{\partial \tau_j}
=& 
\Bigl[\text{diag}\left(\frac{1}{\mu_r},\frac{1}{\mu_t}\right)\Bigr]_{ij} \\
&-\delta_{ij}\Bigl(\frac{m(\tau)}{\tau^2}\Bigr)
- \tau_i\,\tau_j\Bigl(\frac{m'(\tau)}{\tau^3}-\frac{2\,m(\tau)}{\tau^4}\Bigr),
\end{split}
\label{eq:EDOinversemag}
\end{equation}
where the prime indicates the first derivative, leading to inverse magnification 
\begin{equation}
\begin{split}
    \mu^{-1} &= \left| \left(\frac{1}{\mu_r}-\kappa_{\rm \mu l}(\tau)\right)
    \left(\frac{1}{\mu_t} - \kappa_{\rm \mu l}(\tau)\right) 
    -
    \gamma_{\rm \mu l,t}^{\!2}(\tau)
    \right.
    \\
    &
    \left.
    -
    \left(\frac{1}{\mu_r}-\frac{1}{\mu_t}\right)
    \gamma_{\rm \mu l,t}(\tau)
    \cos2\phi
    \right| \,. 
    \label{eq:EDOinversemag2}
\end{split}
\end{equation}
where $\phi$ is the polar angle of $( \tau_1, \tau_2)$ and
where we have defined 
\begin{equation}
\begin{split}
    \kappa_{\rm \mu l}(\tau) &\equiv {m'(\tau)}/{2\tau}\,,\\ \bar\kappa_{\rm \mu l}(\tau) &\equiv {m(\tau)}/{\tau^2},
    \\
    \gamma_{\rm \mu l,t}(\tau) &\equiv \bar\kappa_{\rm \mu l}(\tau) - \kappa_{\rm \mu l}(\tau),
\end{split} 
\end{equation}
as the local convergence, mean convergence, and tangential shear of the microlens.
Note that Eq.~\eqref{eq:EDOinversemag2} reduces to the expression in \cite{Oguri:2017ock} for a point-like lens ($m(\tau)=1$) and to $(\mu_r\mu_t)^{-1}$ in the absence of a microlens ($m(\tau)=0$), as it should.

We can find the critical curve as the combination $(\tau_1,\tau_2)$ which gives $\mu^{-1}=0$ for a given lens profile $m(\tau)$, and the caustic curve as the corresponding locus of source coordinates ($u_{1}$,$u_{2}$).
For point-like lenses the equation can be analytically solved. In the $\tau_1$-direction ($\cos2\phi = 1$), this gives $\tau_1 \sim \sqrt{\mu_t}$, which was found in \cite{Oguri:2017ock} to be a measure of the size of the critical curve. Using the lens equation, this gives the size of the caustic to be $u_1 \sim {\sqrt{\mu_t}}/{\mu_r}$.

For extended lenses, $\mu^{-1}=0$ implies (again for $\cos2\phi = 1$)
\begin{align}\label{eq:EDOcritcurve}
    \tau^2 - \mu_t m(\tau) &= 0 \\
    \text{or} \quad\quad \tau^2 - \mu_r  \left( \tau m'(\tau) - m(\tau)\right) &=0.
    \label{eq:EDOcritcurve2}
\end{align}
The first of these equations corresponds to the point-like critical curve, and since typically $\mu_t \gg \mu_r$, it is usually the outer one. For specific EDO profiles with $\tau m'(\tau) - m(\tau) >0 $ (or $\kappa_{\mu l} > \gamma_{\mu l,t} $, the local convergence of the microlens exceeds its tangential shear), \eqref{eq:EDOcritcurve2} gives rise to extra critical curves and corresponding caustics -- this is a possibility we will discuss in the next section.

In analogy with the point-like lens case, one can find a measure of the size of the critical curve for a specific EDO mass profile by solving equation \eqref{eq:EDOcritcurve}. 
In particular, an EDO acts approximately as a point-like lens if~\cite{Bringmann:2025cht}
\begin{equation}
\tau_m \lesssim \sqrt{\mu_t}, 
\qquad \text{where} \qquad
\tau_m \equiv \frac{\theta_{\rm lens}}{\theta_E} = \frac{R_{\rm lens}}{R_E}.
\end{equation} 
This criterion has an important implication: caustic-crossing events in giant arcs can probe EDOs of much larger 
spatial extent than those accessible to galactic microlensing 
(e.g.~\cite{Croon:2020wpr,Bai:2020jfm,Croon:2020ouk,Croon:2024jhd}, see also \cite{Yang:2025yej}). 
There are two reasons for this. 
First, the (point-like) Einstein radius,
\begin{equation}
R_E = \theta_E D_{\rm L} =  \sqrt{\frac{4GM}{c^2} \frac{D_{\rm L} D_{\rm L S}}{D_{\rm S}} },
\end{equation}
grows with the square root of the lensing distances, which are cosmological for giant arcs. 
Second, the presence of the macrolens provides an additional boost, effectively enlarging the range of EDO sizes that behave like point lenses by a factor of $\sqrt{\mu_t}$. 
Intuitively, this arises because the macrolens stretches the image in the tangential direction, 
so that the effective resolution of the microlens is enhanced.
We may thus as in \cite{Diego:2017drh} define the effective Einstein radius of microlenses in a macro-lensing environment, 
\begin{equation}
\bar{R}_E \equiv \sqrt{\frac{4GM \mu_t}{c^2} \frac{D_{\rm L} D_{\rm L S}}{D_{\rm S}} },
\label{eq:effRE}
\end{equation}
which is relevant for all types of microlenses. As we will see in Section~\ref{sec:eff}, we may further define an effective Einstein radius for extended lenses by means of an efficiency with respect to point-like lenses, which will be useful in generalising constraints on DM.

\section{EDO light curves}

\begin{figure*}
    \centering
    \includegraphics[width=.49\linewidth]{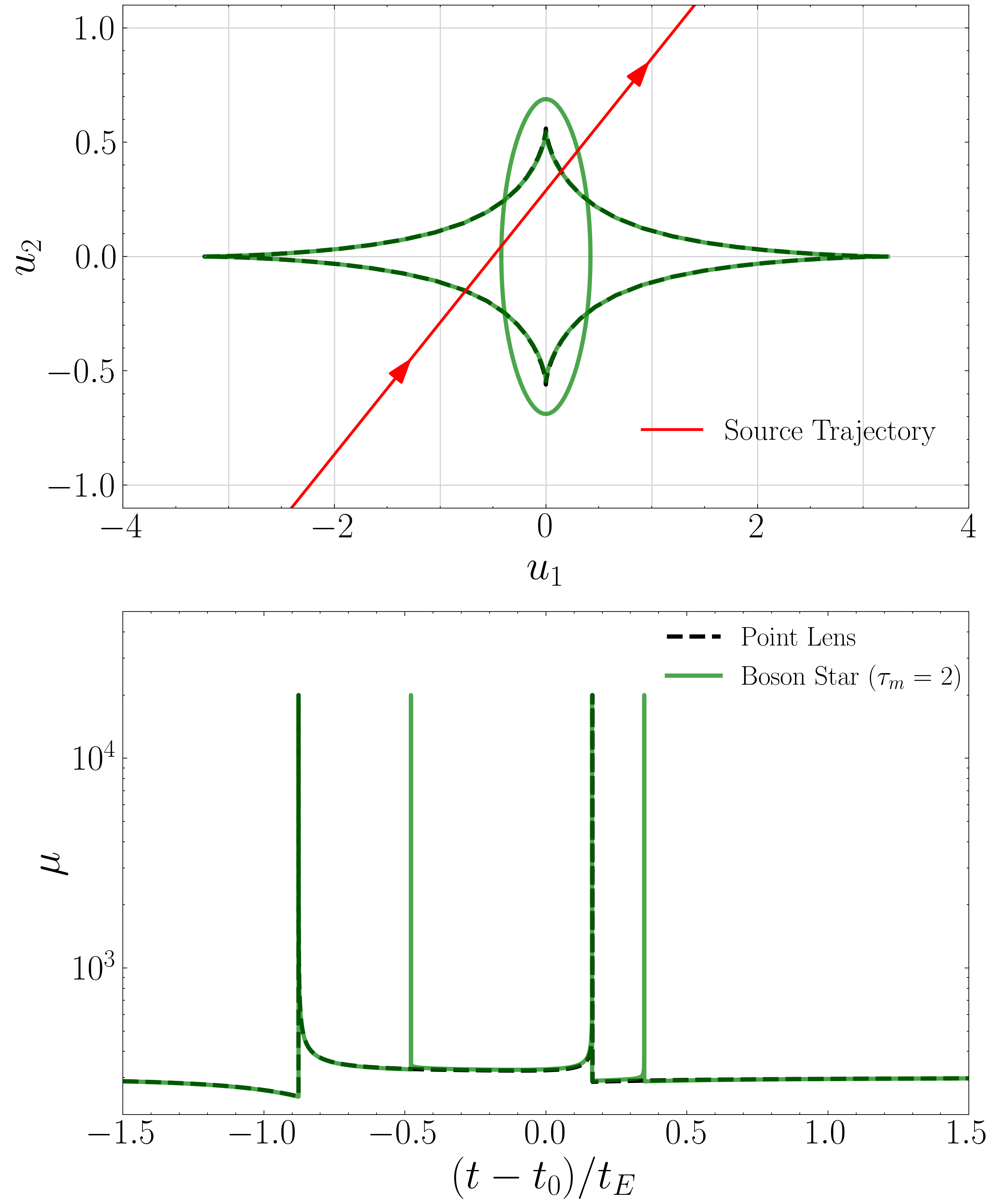}
    \includegraphics[width=.49\linewidth]{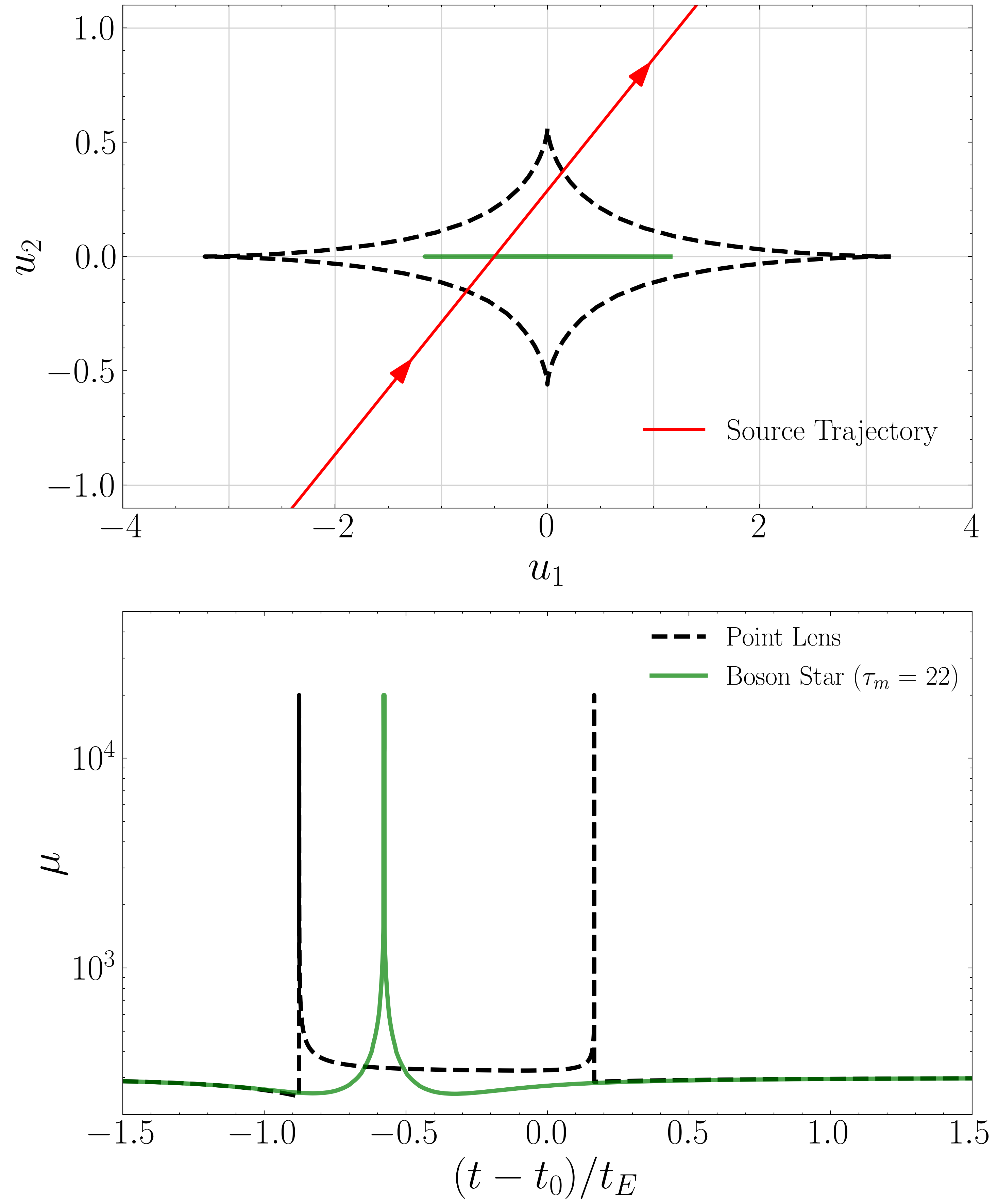}
    \caption{Boson star caustics compared to point-like lens caustics for a macrolens with $\mu_{\rm{t}} \sim 100$ and $\mu_{\rm{r}}\sim3$, and light curves resulting from the indicated source trajectory. \textit{Left:} $\tau_m=2$ gives rise to extra caustics in the light curve; 
    \textit{Right:} for $\tau_m=22$ ($\gtrsim \sqrt{\mu_t}$) the caustics are spaced so closely together that they appear as one in the light curve.
    }
    \label{fig:tau_m_2_22_plots}
\end{figure*}

In this section we study the circumstances under which the microlensing light curves due to EDO lenses deviate from those produced by point-like lenses.
\vtwo{We consider the specific case of a boson star~\cite{JETZER1992163,Dabrowski:1998ac}: a gravitationally bound object comprised of elementary or composite scalar fields, with kinetic pressure 
preventing gravitational collapse. We derive a lens-plane-projected mass profile $m(\tau)$ as in~\cite{Croon:2020wpr}, by numerically solving the Schr\"odinger-Poisson equations that describe their hydrostatic equilibrium in the non-relativistic limit and under the assumption of negligible self-coupling.}
Using the analytic model in the previous section, we calculate the magnification as a function of time from a given source trajectory ($u_{1}(t)$,$u_{2}(t)$) by solving equations~\ref{eq:lensingeqns} for the image positions $(\tau_{1}(t),\tau_{2}(t))$ and plugging the results into~\ref{eq:EDOinversemag2}. Summing over all images gives us the light curve.
The caustic curve is found by equating \eqref{eq:EDOinversemag2} with zero.

For a point-like microlens, the number of images depends on whether the source lies inside or outside the caustic diamond. Outside the diamond the lens produces two images, while inside it produces four. As the source approaches the caustic from the outside, no additional images are created and the total magnification remains smooth and finite. In contrast, when approaching the caustic from the inside, the extra pair of images moves toward the critical curve; their Jacobian determinants vanish and (in the point-source approximation) their magnifications diverge as $\mu \propto d^{-1/2}$, where $d$ is the perpendicular distance to the caustic. 
Thus the divergence of the magnification is intrinsically one-sided, reflecting that fold caustics correspond to the birth or annihilation of an image pair rather than a symmetric enhancement on both sides. This behaviour is demonstrated for a randomly chosen trajectory, and assuming the same tangential and radial macro magnifications found for MACS J1149 LS1 in~\cite{Oguri:2017ock}, $\mu_{\rm{t}} \sim 100$ and $\mu_{\rm{r}}\sim3$, in Fig.~\ref{fig:tau_m_2_22_plots}.
The caustic curve is presented in the source plane with origin aligned with the lens centre, and the source trajectory is indicated. This is shown alongside the corresponding light curve, with time given in units of the Einstein crossing time $t_{\rm E} = D_{\rm L}\,\theta_{\rm E}/{v_\perp}$. Here we define $t_{0}$ as the time at which the source crosses the vertical axis $u_{2}$.

\vtwo{In the same figure, we compare this to an extended lens. As mentioned previously, we have chosen a lens with a boson star profile as an illustrative example, since its shallow core
($\kappa_{\rm \mu l} \simeq \bar\kappa_{\rm \mu l}$) suppresses the tangential shear and produces light curves that deviate most strongly from the point-lens case.}
For a range in $\tau_m \equiv  \theta_{\rm lens}/\theta_E =R_{\rm lens}/R_E $ (where here we adopt $R_{\rm lens} = R_{90}$ as a characteristic size: the radial coordinate which encloses $90\%$ of the objects total mass), we find that, in addition to the diamond-shaped caustic curve of a point-like lens, a secondary caustic curve appears. 
At very large impact parameters, when the source is far from both caustics in the source plane, this boson star case features one image. This image closely mirrors the point-like image which carries all of the flux. As the source crosses the outer caustics, two images are created, and upon entering the inner caustics, a further two images are created, giving a total of five. Pairs of images are annihilated again as the source leaves the area enclosed by the caustics in the source plane.
Such a scenario leads to two additional caustic features in the light curve. 
We demonstrate this for  $\tau_m = 2$ in the left panel of Fig.~\ref{fig:tau_m_2_22_plots}.
Depending on the source trajectory, these extra features may occur at larger or smaller values of $t-t_0$ compared to the point-like case. They are, however, typically very narrow in time, which makes them observationally challenging. For comparison, the Icarus star caustic crossing was observed on the timescale of weeks, suggesting that such features might be accessible with sufficiently regular cadence. It is also important to note that these solutions assume a point-like source, and finite source effects could further modify the signal. 

For somewhat larger boson stars ($\tau_m \sim 22$), 
a single caustic curve is found which is squeezed in the $u_2$ direction.
Thus, the number of caustic crossings is the same as in the point-like case, but their appearance in the light curve changes: realistic trajectories produce what appears to be a single symmetric caustic rather than two. This is shown in the right panel of Fig.~\ref{fig:tau_m_2_22_plots}. Finally, when the boson star radius exceeds $R_{90}\gg\sqrt{\mu_t}\,R_E$ (corresponding to $\tau_m \gg \sqrt{\mu_t}$), the caustics disappear entirely, as expected. 

\vtwo{We can conclude that the light curves of caustic-crossing stars involving the presence of extended microlenses can include qualitative differences to those involving point-like microlenses, which are dependent on their internal mass profiles and their physical scale. In practice, however, distinguishing between point-like and extended microlenses will unlikely be possible by observing any single event.
Stochastic aspects of the caustic network and source trajectory imply that the detailed light curve structure of any real such event cannot be predicted deterministically. 
Consequently, separating point-like from extended microlenses is only feasible statistically, requiring ensembles of well-characterised caustic crossing events spanning a range of microlens parameters, macro-magnifications and source trajectories, together with sufficient cadence and signal-to-noise to resolve short-timescale features and mitigate degeneracies with the macrolens model.
}

\section{Constraining EDOs}
\subsection{Extended lens efficiency}\label{sec:eff}

\begin{figure}[t]
    \centering
    \includegraphics[width=0.95\linewidth]{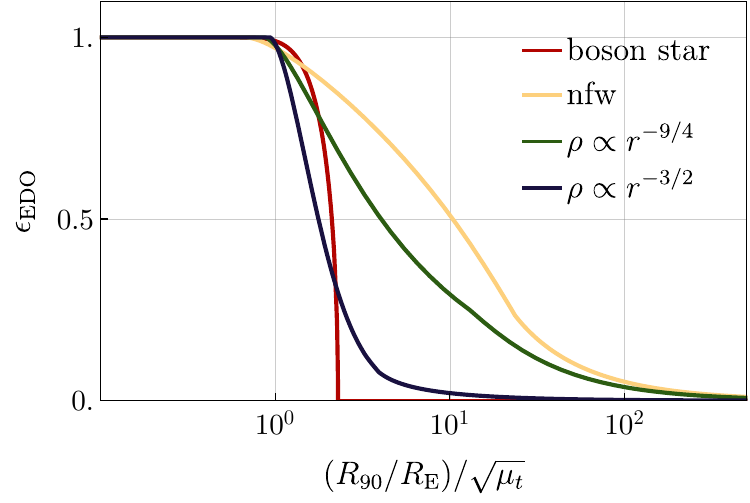}
    \caption{Extended lens efficiency for different examples of EDOs (see \cite{Croon:2020wpr} for further details). As expected, for $R_{90}\leq \sqrt{\mu_t} R_{\rm E}$, EDOs are effectively point-like, whereas they become less efficient when they are more dilute. 
    }
    \label{fig:efflens}
\end{figure}

In Ref.~\cite{Croon:2020wpr}, the concept of a microlensing efficiency for extended lenses was introduced, which will be useful for our purposes. 
This efficiency captures the effective reduction of the Einstein radius of an extended lens with respect to a point-like lens.
By analogy, we define an \emph{effective Einstein radius} for an extended lens in a macrolensing environment as
\begin{equation}
    \bar{R}_{E,\rm EDO} \equiv \epsilon_{\rm EDO} \, \sqrt{\mu_t} \, \theta_E D_L,
    \label{eq:efficiencies}
\end{equation}
where the efficiency factor $\epsilon_{\rm EDO}$ is obtained from the normalized form of 
Eq.~\eqref{eq:EDOcritcurve},
\begin{equation}
    \epsilon_{\rm EDO}^2 - m\!\left(\epsilon_{\rm EDO}\sqrt{\mu_t}\right) = 0.
\end{equation}
For a point-like lens, $\epsilon_{\rm EDO}=1$, and the effective Einstein radius is simply given by the size 
of the point-like critical curve, i.e.
$$
\tau = \sqrt{\mu_t}, \qquad 
\theta = \theta_E \sqrt{\mu_t},
$$
in agreement with Eq.~(9) of Ref.~\cite{Oguri:2017ock} and Ref.~\cite{Diego:2017drh}. 
Intuitively, $\epsilon_{\rm EDO}$ quantifies the reduction in lensing strength due to the extended structure 
of the object: a more diffuse mass distribution produces weaker deflections near the center, effectively 
shrinking the size of the region over which the lens can generate appreciable magnification. 
Figure~\ref{fig:efflens} illustrates this construction for several example EDO profiles introduced in Ref.~\cite{Croon:2020wpr}.

We note that our definition of efficiency is conservative, because we focus on the point-like critical curves \eqref{eq:EDOcritcurve} and corresponding caustics. As seen from \eqref{eq:EDOcritcurve2} and demonstrated explicitly in the previous section, EDO microlensing events may feature additional high-magnification features in the light curve which are not captured by this definition, but are important for larger radii.

\subsection{Icarus constraint}

In Ref.~\cite{Oguri:2017ock}, a constraint on the fraction of point-like objects was derived from the 
MACS~J1149 LS1 (``Icarus'') event. 
The basic idea is that even a small abundance of point-mass lenses can dramatically alter the asymptotic 
behaviour of the macro-model magnification near a critical curve. 
Because the \emph{effective} Einstein radius \eqref{eq:effRE} of each microlens grows as $\sqrt{\mu_t}$, 
these radii begin to overlap at large $\mu_t$ even for modest lens surface densities \cite{Diego:2017drh}. 
Once this ``saturation'' threshold is reached, ray-tracing simulations show 
that the source is split into a large number of micro-images, 
and the total magnification becomes insensitive to the exact position of the source 
relative to the macromodel caustic.  
The peak brightness of the Icarus event can then be used to constrain the required magnification and therefore the maximum fraction of mass in  the form of microlenses.

Let us first review the minimum peak magnitude in the MACS J1149 LS1 Icarus event calculated in Ref.~\cite{Oguri:2017ock}.
In caustic crossing events in giant arcs, the magnification is limited by finite-source effects. 
The largest possible magnification occurs once the distance to the caustic becomes comparable to the source size in the source plane
$\beta_R = R/D_S$.
For point lenses, both numerical simulations and analytic estimates give
\begin{equation}
\mu_{\rm max} \;\approx\; \mu_t \mu_r \sqrt{\frac{\theta_E}{\sqrt{\mu_t}\,\beta_R}} \, .
\label{eq:mumax}
\end{equation}
We note that this is an idealised equation, and deviations may occur at the $\mathcal{O}(1)$ level~\cite{Oguri:2017ock}.
This maximum magnification can be converted into an apparent peak magnitude by assuming a stellar temperature 
and bolometric correction~\cite{Oguri:2017ock},
\begin{equation}
a_{\rm peak}
\;\approx\; 46.4 - 5 \log_{10} \!\left(\frac{R_{\rm source}}{R_\odot}\right) 
- 2.5 \log_{10}\mu_{\rm max} ,
\end{equation}
\vtwo{for which we have used $a$ to represent the apparent magnitude instead of the conventional $m$ to avoid confusion with the normalised lens-plane-projected mass.}

An independent constraint on the source radius comes from the source-crossing time, 
$t_{\rm src} = 2 R_{\rm source} / v (1+z_s)$, with $v$ the effective transverse velocity in the source plane and $z_s = 1.49$ the redshift of the source. 
Since the light curve exhibits finite-source effects only on timescales shorter than $\sim 10$ days, 
one infers $R_{\rm source} \lesssim 260\,R_\odot$ for $v=500\,\mathrm{km\,s^{-1}}$~\cite{Oguri:2017ock}. 
One can then find a minimum peak magnitude using this upper bound on the source radius.
Substituting in this the upper bound and Eq.~\eqref{eq:mumax} with $\mu_r=3$, $\theta_E = 1.8 \times 10^{-6} (M/M_\odot)^{1/2}\,\rm arcsec $, $ D_L = 6.4 \, \rm kpc \, arcsec^{-1}$, and $ D_S = 8.5 \, \rm kpc \, arcsec^{-1}$~\cite{Oguri:2017ock} yields
\begin{align}
\begin{split}
    a_{\rm peak, min}
    &\;\approx\; 25.1 - 0.625 \log_{10}\!\left(\frac{M}{M_\odot}\right) 
    \\
    &\qquad - 1.875 \log_{10}\!\left(\frac{\mu_t}{100}\right),
\end{split}
\end{align}
to be compared with the observed Icarus peak magnitude $a_{\rm peak} \sim 26$. In this equation, we remind the reader that $M$ is the mass of the microlens, which enters through $\mu_{\rm max}$. 

\begin{figure}
    \centering
    \vspace{-7mm}
    \includegraphics[width=1.0\linewidth]{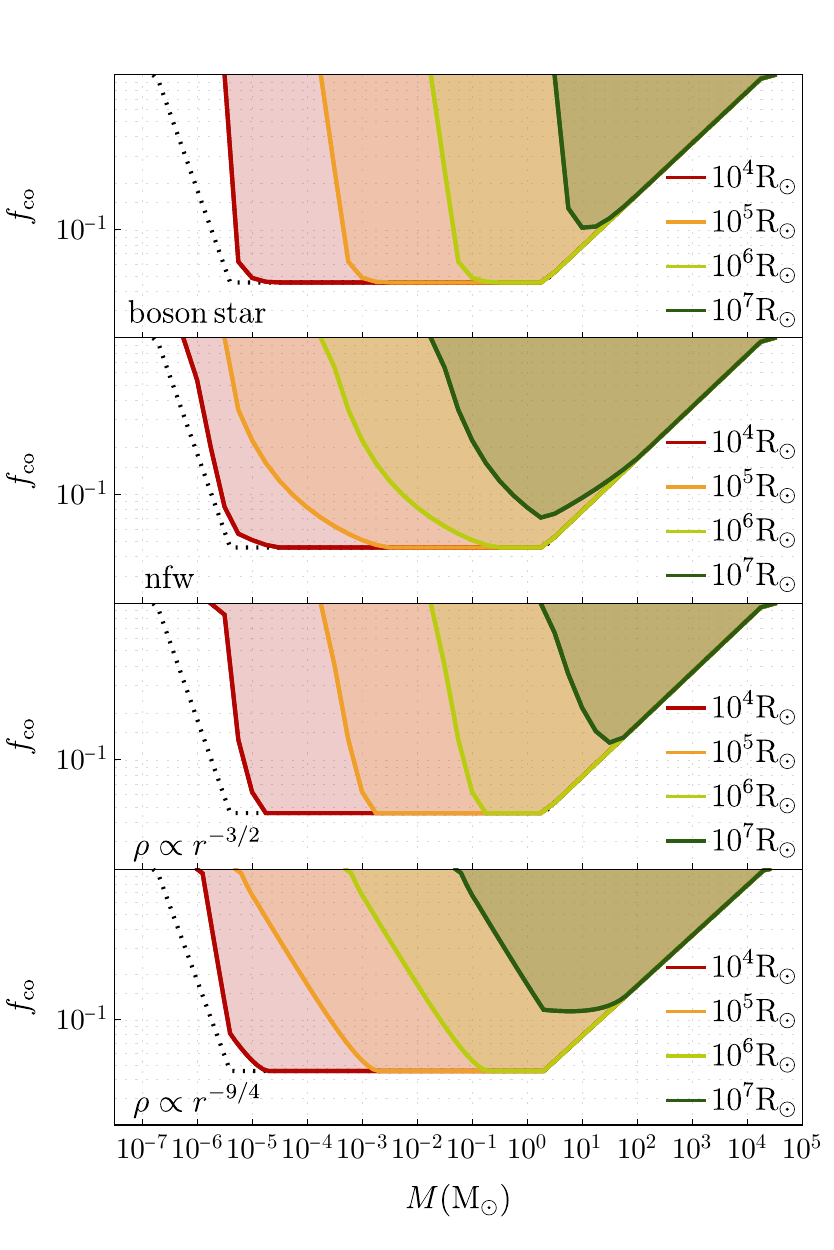}
    \caption{Constraints on the compact object fraction $f_{\rm co}$ from the MACS J1149 LS1 Icarus event. Colored lines show the constraints for extended dark objects (EDOs) of different radii while the dotted line shows the constraints for point-like objects such as primordial black holes. As shown in text, we expect $f_{\rm DM} \simeq [0.92-0.97] f_{\rm co} < f_{\rm co}$ for this event.
    }
    \label{fig:constraints}
\end{figure}

The presence of substructure reduces the high magnification near the critical curve, and transfers it to regions further away from it. 
The more substructure near the critical curve, the smaller the peak magnification. Then, through the observation of the MACS J1149 LS1 Icarus event, this mechanism can be used to derive constraints on the amount of substructure \cite{Venumadhav:2017pps,Diego:2017drh,Oguri:2017ock,Palencia:2023kne}.               
The effect can be quantified through an optical depth,
\begin{equation}
    \Theta = \frac{\Sigma}{M}\,\pi \bar{R}_E^2 ,
    \label{eq:optdepthPL}
\end{equation}
where $\Sigma$ is the surface mass density of lenses, $M$ is the mass of each lens.
Saturation occurs when $\Theta\sim1$ \cite{Diego:2017drh},  
which sets a saturation tangential magnification
\begin{equation}
    \mu_{t,\rm sat} = \frac{M}{\pi \Sigma R_E^2}\, .
    \label{eq:mutmax}
\end{equation}
It is particularly convenient to express $\Sigma$ in terms of the fraction of matter in compact objects which act as microlenses, 
\begin{equation}
    f_{\rm co} \equiv \frac{\Sigma}{\Sigma_{\rm crit}} \kappa^{-1},
\end{equation}
where $ \kappa \equiv \Sigma_{\rm tot}/\Sigma_{\rm crit} = 0.83$ is the (total) surface mass density, and where 
$\Sigma_{\rm crit} = M/(\pi R_E^2)=2.4 \times 10^3\,M_\odot\,\mathrm{pc}^{-2}$ 
for $\theta_E$ and $D_L$ as before~\cite{Oguri:2017ock}. 
In these units, the saturation condition becomes
\begin{equation}
    \mu_{t,\rm sat} =  \kappa^{-1} f_{\rm co}^{-1}.
    \label{eq:mutmax2}
\end{equation}
For extended dark objects (EDOs), we may use our concept of an effective EDO Einstein radius in a macrolensing environment \eqref{eq:efficiencies} 
to find an EDO optical depth
\begin{equation}
    \Theta_{\rm EDO} = \frac{\Sigma}{M}\,\pi \bar{R}_{E,\rm EDO}^2 = \frac{\Sigma}{M}\,\pi \epsilon_{\rm EDO}^2 \mu_t R_E^2 .
    \label{eq:opticaldepth2}
\end{equation}
The maximum tangential magnification for EDOs is obtained by setting $\Theta_{\rm EDO}=1$, leading to 
\begin{equation}
    \mu_{t,\rm sat, EDO} \simeq \kappa^{-1} f_{\rm co}^{-1} \epsilon_{\rm EDO}^{-2}.
\end{equation}
Using the saturation tangential magnification one can then constrain the compact object fraction from above by equating
\begin{equation}
     a_{\rm peak, min} \biggr\rvert_{\mu_t = \mu_{t, \rm sat}} \approx 26.
\end{equation} 
To set these constraints, we assume a monochromatic mass distribution of DM objects, noting that constraints can be generalised straightforwardly (e.g. see \cite{Carr:2017jsz,Sevi2024EDObounds}). We also need to make an assumption for the mass of the microlens which explains the observation.
We distinguish three regimes:
\begin{itemize}
    \item For DM objects of mass $< 2 \,M_\odot $, we make the conservative assumption that the Icarus event is explained not by a DM lens but by an intracluster (ICL) star of $M=2\,M_\odot$.\footnote{This assumed ICL mass is conservative because typical ICL stars have much smaller masses (in the range $0.1-1 \rm \, M_\odot$), which would lead to stronger constraints as the needed $\mu_t$ is larger. The assumption also affects the size of the objects which can be constrained through the $\mu_t$-dependent efficiency.}
    \item For heavier lenses than the assumed ICL star ($M \geq 2\,M_\odot$), a DM lens is itself invoked to explain the observation.
    This leads to a constraint which becomes weaker with increasing mass, since the required $\mu_t$ is smaller for a larger lens mass. 
    \item At very low lens masses, the microcaustic size becomes smaller than the source. This implies that the lensing signal is smeared out and the maximum magnification cannot be reached. Constraints are therefore no longer applicable when 
    \begin{equation} \label{eq:lowerM}
    \frac{\theta_E}{\sqrt\mu_t} < \beta_R ,
    \end{equation}
    with $\beta_R = 2.7 \times 10^{-12} \,(R_{\rm source}/R_{\odot})\, \mathrm{arcsec}$ being the angular size of the source, and $R_{\rm source}$ its physical size in the source plane subject to the bound above. In Ref.~\cite{Oguri:2010rh}, the bound $\mu_t <100$ was used to find a lower $M$ for which the constraints were valid; here we impose the constraint \eqref{eq:lowerM} directly. 
\end{itemize}
This procedure leads to the constraints shown in Fig.~\ref{fig:constraints}.
In these plots, the dotted lines correspond to the constraints on point-like microlenses (including PBHs) and can be compared directly to the results in Ref.~\cite{Oguri:2017ock}. The difference is primarily explained by the assumed ICL lens mass, taken to be $10 M_\odot$ in \cite{Oguri:2017ock}. We note here that such heavy stars have lifetimes of only tens of Myr (see e.g.~Ref.~\cite{Spera:2015vkd}), far below that of observed ICL stars~\cite{DeMaio:2015gka}.

We also note that the radii of the objects constrained in Fig.~\ref{fig:constraints} are much larger than can be constrained through conventional microlensing observations \cite{Croon:2020wpr,Croon:2020ouk}. This is a result of both the cosmological distances involved and the $\sqrt{\mu_t}$ boost in the Einstein radius from the macrolens.

Finally, we connect the \emph{local} compact object fraction at the arc position $f_{\rm co}$ to the cosmological 
fraction of dark matter in compact objects. 
We define
$$
f_{\rm DM} \equiv \frac{\rho_{\rm DO}}{\rho_{\rm DM}} .
$$
Along the relevant line of sight, 
$$
\Sigma_{\rm co} = \Sigma_* + \Sigma_{\rm DO},
$$
where $\Sigma_*$ is the surface density in ICL stars and remnants, and 
$\Sigma_{\rm DO}$ is the compact dark object contribution. 
If compact dark objects trace the local DM sheet, then 
$\Sigma_{\rm DO} = f_{\rm DM}\,\Sigma_{\rm DM}$, giving
\begin{equation}
f_{\rm co} = \frac{\Sigma_*}{\Sigma_{\rm tot}} 
+ f_{\rm DM}\,\frac{\Sigma_{\rm DM}}{\Sigma_{\rm tot}}
\;\equiv\; f_* + f_{\rm DM}\,s,
\end{equation}
with $f_* \equiv \Sigma_*/\Sigma_{\rm tot}$ 
and $s \equiv \Sigma_{\rm DM}/\Sigma_{\rm tot}$ the local dark-matter fraction at the arc. 
Thus,
\begin{equation}
f_{\rm DM} = \frac{f_{\rm co} - f_*}{s}\, .
\end{equation}
For ``clean'' giant arcs that avoid bright galaxies, 
one expects $0.003 \lesssim f_* \lesssim 0.01$ (including the $f_* \sim 0.007$ value of 
Ref.~\cite{Oguri:2017ock}) and $s\simeq 0.94$--$0.99$, 
such that $f_{\rm DM} \simeq [0.92-0.97] f_{\rm co} < f_{\rm co}$.

\section{Outlook}
The constraint derived from the Icarus event is presently limited by the uncertainty in the tangential magnification $\mu_t$ at the location of the star. 
Since $\mu_t$ is not a directly observable quantity but instead depends on details of the cluster lens model, the resulting limits on $f_{\rm co}$ were derived parametrically in $\mu_t$. 
The robustness and competitiveness of the constraint therefore depends on how well one can model the local shear and convergence near the cluster caustic.

Because the saturation threshold scales as $\mu_t^{-1}$, 
systems with large and accurately known $\mu_t$ will exclude even tiny abundances of compact objects, as long as the event is not washed out by finite-source effects. 
Moreover, since the effective Einstein radius scales like $\bar{R}_E \propto \sqrt{\mu_t}$, an accurately measured and large value $\mu_t \sim 10^4$ would boost sensitivity to lenses a factor $ \sim 10$ larger in radius.  
Thus, future detections in high-magnification arcs could lead to constraints on DM fractions that are substantially tighter and less model-dependent than those obtained from Icarus.
Several avenues are particularly promising:
\begin{itemize}[noitemsep,leftmargin=1.em]
    \item \textbf{Giant arcs with well-constrained lens models.} 
    If a caustic-crossing star is discovered in a bright arc containing many multiply imaged knots (such as the Dragon arc in Abell 370), the critical curve can be pinned down with much higher accuracy than in MACS~J1149. 
    This would reduce the uncertainty on $\mu_t$ from orders of magnitude to the tens-of-percent level.
    \item \textbf{Multiple transients in the same arc.} 
    The detection of several caustic star events along a single arc provides multiple independent probes of the magnification pattern. 
    Joint modeling of such systems would strongly constrain the local shear and convergence.
    \item \textbf{Well-studied cluster lenses.} 
    Clusters such as Abell~370 or MACS~J0416 already have extensive spectroscopy and high-resolution imaging. 
    In these systems, lens models are in general better constrained, 
    and any caustic star events detected there would yield correspondingly stronger compact-object limits.
    
    \item \textbf{Dedicated time-domain surveys.} 
    State-of-the-art facilities like \emph{JWST}, the \emph{Roman Space Telescope}, and \emph{LSST} will 
    monitor strongly lensed fields over long baselines, potentially discovering 
    hundreds of caustic transients. 
    A statistical sample of events, combined with high-fidelity lens models, 
    will allow population-level constraints on compact dark objects that are 
    far more stringent than those available from Icarus alone.
\end{itemize}

Since the observation of MACS~J1149 LS1, other caustic crossings in giant arcs have been observed. The ``Godzilla'' event in the Sunburst Arc ($z=2.37$) reached $\mu \sim 600$ with microcaustic peaks up to a few $10^3$ \cite{Diego:2022mii}; ``Spock'' transients in MACS J0416 showed $\mu \sim 400$ \cite{rodney2018two}; ``Mothra'' may exceed $\mu \sim 6000$ if its compact perturber is confirmed \cite{Diego:2023qhp}. ``Earendel'' (at $z=6.2$) has been found to have $\mu > 4000$ from JWST imaging \cite{welch2022jwst}.\footnote{However, lens models in Ref.~\cite{scofield2025earendel} suggest a lower value of $\mu \approx 43$-$67$. This discrepancy highlights the sensitivity of such constraints to lens modeling systematics and underscores the need for continued development of cluster mass reconstructions.}
These increasingly numerous caustic-crossing stars may offer a pathway to probe smaller compact objects than previously accessible. 
Perhaps one of the most promising galaxies to improve on these constraints is the Dragon arc at $z=0.725$ where over 40 microlensing events have been observed with JWST data \cite{Fudamoto:2024idh}. Already granted new observations in JWST's cycles 5, 6 and 7 will increase this number even further providing a unique statistical data set of microlensing events from a single galaxy.

Moreover, complementary work \cite{Dai:2019lud} has demonstrated that a population of low mass, diffuse axion minihalos can collectively produce surface-density fluctuations at the level of $\Delta\kappa \sim 10^{-4}$--$10^{-3}$, imprinting small-scale irregularities in the light curve. These stochastic perturbations from a dense ensemble of subhalos probe a different regime of the same parameter space from the one explored in this work. In principle, detecting the absence of such irregularities in well-sampled light curves could provide complementary limits on extremely low mass ($M \sim 10^{-15}$--$10^{-8}\,M_\odot$) and diffuse dark objects. 

\section{Conclusion}
In this work we have extended the study of caustic-crossing stars in giant arcs to the case of extended dark objects. By developing an analytic framework for microlensing in a cluster environment, we have shown how the finite size of the lens modifies both the critical curve structure and the resulting light curves. Depending on their radius relative to the Einstein radius, EDOs can act as effectively point-like lenses, generate additional narrow caustics, effectively merge caustics, or wash out the caustic structure entirely. We quantified the resulting reduction in lensing efficiency through an effective Einstein radius, and demonstrated how this generalises existing treatments of point lenses. Applying this formalism to the MACS J1149 LS1 Icarus event, we derived constraints on the fraction of dark matter that may reside in extended objects. 

Our results highlight that caustic crossings in giant arcs probe a complementary region of parameter space compared to traditional galactic microlensing surveys. In particular, the combination of cosmological lensing distances and the strong tangential magnification of the cluster macrolens extends sensitivity to dark objects with $ R \lesssim 10^7 \rm R_\odot$ -- much larger radii than can be accessed in the Galaxy. 
This has important implications for cosmology, as ultracompact minihaloes formed from the collapse of primordial overdensities can be probed \cite{Bringmann:2025cht}.
The increasing number of observed caustic-crossing stars, combined with upcoming facilities such as JWST, Roman, and LSST, promises to transform this field from isolated case studies into a statistical probe of dark matter structure. With improved lens models and a growing sample of transients, future observations will be able to place significantly sharper and more robust constraints on the abundance of compact and extended dark objects across a wide mass and size range.

\section*{Software}
Wolfram Mathematica version 14.2.1, python 3.13.7.

\section*{Acknowledgements}
DC and BC are supported by the STFC under Grant No.~ST/T001011/1. DC is grateful to the Mainz Institute for Theoretical Physics (MITP) of the Cluster of Excellence PRISMA+ (Project ID 390831469) for its hospitality and its partial support during the completion of this work.
JMD acknowledges the support of projects PID2022-138896NB-C51 (MCIU/AEI/MINECO/FEDER, UE) Ministerio de Ciencia, Investigaci\'on y Universidades. 
BJK acknowledges support from the \textit{Consolidaci\'on Investigadora} Project \textsc{DarkSpikesGW}, reference CNS2023-144071, financed by MCIN/AEI/10.13039/501100011033 and by the European Union ``NextGenerationEU"/PRTR. BJK also acknowledges support from the project \textsc{DMpheno2lab} (PID2022-139494NB-I00) financed by MCIN/AEI/ 10.13039/501100011033/FEDER, EU.
BJK and JMD also acknowledge support from the project SA101P24 (Junta de Castilla y León).
JMP acknowledges financial support from the Formaci\'on de Personal Investigador (FPI) programme, ref. PRE2020-096261, and partial finantial support from the Complementary Plan in Astrophysics and High-Energy Physics (CA25944), project C17.I02.P02.S01.S03 CSIC, supported by the Next Generation EU funds, RRF and PRTR mechanisms, and the Government of the Autonomous Community of Cantabria.

\bibliography{refs}

\end{document}

%% file: universalnewcommands.tex
\newcommand{\beq}{\begin{equation}}
\newcommand{\eeq}{\end{equation}}
\newcommand{\bea}{\begin{eqnarray}}
\newcommand{\eea}{\end{eqnarray}}

\usepackage{pdfbase}[2017/03/16]
\usepackage{xparse,ocgbase}
\usepackage{xcolor,calc}
\usepackage{tikzpagenodes,linegoal}
\usetikzlibrary{calc}
\usepackage{tcolorbox}

\ExplSyntaxOn
\let\tpPdfLink\pbs_pdflink:nn
\let\tpPdfAnnot\pbs_pdfannot:nnnn\let\tpPdfLastAnn\pbs_pdflastann:
\let\tpAppendToFields\pbs_appendtofields:n
\def\tpPdfXform{\pbs_pdfxform:nnnnn{1}{1}{}{}}
\let\tpPdfLastXform\pbs_pdflastxform:
\let\cListSet\clist_set:Nn\let\cListItem\clist_item:Nn
\ExplSyntaxOff

\usepackage{pdfbase}[2017/03/16]
\usepackage{xparse,ocgbase}
\usepackage{xcolor,calc}
\usepackage{tikzpagenodes,linegoal}
\usetikzlibrary{calc}
\usepackage{tcolorbox}

\ExplSyntaxOn
\let\tpPdfLink\pbs_pdflink:nn
\let\tpPdfAnnot\pbs_pdfannot:nnnn\let\tpPdfLastAnn\pbs_pdflastann:
\let\tpAppendToFields\pbs_appendtofields:n
\def\tpPdfXform{\pbs_pdfxform:nnnnn{1}{1}{}{}}
\let\tpPdfLastXform\pbs_pdflastxform:
\let\cListSet\clist_set:Nn\let\cListItem\clist_item:Nn
\ExplSyntaxOff

\makeatletter
\NewDocumentCommand{\tooltip}{%
  ssssO{\ifdefined\@linkcolor\@linkcolor\else blue\fi}mO{yellow!20}mO{0pt,0pt}%
}{{%
  \leavevmode%
  \IfBooleanT{#2}{%
    \ocgbase@new@ocg{tipOCG.\thetcnt}{%
      /Print<</PrintState/OFF>>/Export<</ExportState/OFF>>%
    }{false}%
    \xdef\tpTipOcg{\ocgbase@last@ocg}%
    \ocgbase@add@ocg@to@radiobtn@grp{tool@tips}{\ocgbase@last@ocg}%
  }%
  \tpPdfLink{%
    \IfBooleanTF{#4}{%
      /Subtype/Link/Border[0 0 0]/A <</S/SetOCGState/State [/Toggle \tpTipOcg]>>
    }{%
      /Subtype/Screen%
      /AA<<%
        \IfBooleanTF{#3}{%
          /E<</S/SetOCGState/State [/Toggle \tpTipOcg]>>%
        }{%
          \IfBooleanTF{#2}{%
            /E<</S/SetOCGState/State [/ON \tpTipOcg]>>%
            /X<</S/SetOCGState/State [/OFF \tpTipOcg]>>%
          }{
            \IfBooleanTF{#1}{%
              /E<</S/JavaScript/JS(%
                var fd=this.getField('tip.\thetcnt');%
                if(typeof(click\thetcnt)=='undefined'){%
                  var click\thetcnt=false;%
                  var fdor\thetcnt=fd.rect;var dragging\thetcnt=false;%
                }%
                if(fd.display==display.hidden){%
                  fd.delay=true;fd.display=display.visible;fd.delay=false;%
                }else{%
                  if(!click\thetcnt&&!dragging\thetcnt){fd.display=display.hidden;}%
                  if(!dragging\thetcnt){click\thetcnt=false;}%
                }%
                this.dirty=false;%
              )>>%
            }{%
              /E<</S/JavaScript/JS(%
                var fd=this.getField('tip.\thetcnt');%
                if(typeof(click\thetcnt)=='undefined'){%
                  var click\thetcnt=false;%
                  var fdor\thetcnt=fd.rect;var dragging\thetcnt=false;%
                }%
                if(fd.display==display.hidden){%
                  fd.delay=true;fd.display=display.visible;fd.delay=false;%
                }%
               this.dirty=false;%
              )>>%
              /X<</S/JavaScript/JS(%
                if(!click\thetcnt&&!dragging\thetcnt){fd.display=display.hidden;}%
                if(!dragging\thetcnt){click\thetcnt=false;}%
                this.dirty=false;%
              )>>%
            }%
            /U<</S/JavaScript/JS(click\thetcnt=true;this.dirty=false;)>>%
            /PC<</S/JavaScript/JS (%
              var fd=this.getField('tip.\thetcnt');%
              try{fd.rect=fdor\thetcnt;}catch(e){}%
              fd.display=display.hidden;this.dirty=false;%
            )>>%
            /PO<</S/JavaScript/JS(this.dirty=false;)>>%
          }%
        }%
      >>%
    }%
  }{{\color{#5}#6}}%
  \sbox\tiptext{%
    \IfBooleanT{#2}{%
      \ocgbase@oc@bdc{\tpTipOcg}\ocgbase@open@stack@push{\tpTipOcg}}%
    \tcbox[colframe=black,colback=#7,size=fbox,arc=1ex,sharp corners=southwest]{#8}%
    \IfBooleanT{#2}{\ocgbase@oc@emc\ocgbase@open@stack@pop\tpNull}%
  }%
  \cListSet\tpOffsets{#9}%
  \edef\twd{\the\wd\tiptext}%
  \edef\tht{\the\ht\tiptext}%
  \edef\tdp{\the\dp\tiptext}%
  \tipshift=0pt%
  \IfBooleanTF{#2}{%
    \setlength\whatsleft{\linegoal}%
  }{%
    \measureremainder{\whatsleft}%
  }%
  \ifdim\whatsleft<\dimexpr\twd+\cListItem\tpOffsets{1}\relax%
    \setlength\tipshift{\whatsleft-\twd-\cListItem\tpOffsets{1}}\fi%
  \IfBooleanF{#2}{\tpPdfXform{\tiptext}}%
  \raisebox{\heightof{#6}+\tdp+\cListItem\tpOffsets{2}}[0pt][0pt]{%
    \makebox[0pt][l]{\hspace{\dimexpr\tipshift+\cListItem\tpOffsets{1}\relax}%
    \IfBooleanTF{#2}{\usebox{\tiptext}}{%
      \tpPdfAnnot{\twd}{\tht}{\tdp}{%
        /Subtype/Widget/FT/Btn/T (tip.\thetcnt)%
        /AP<</N \tpPdfLastXform>>%
        /MK<</TP 1/I \tpPdfLastXform/IF<</S/A/FB true/A [0.0 0.0]>>>>%
        /Ff 65536/F 3%
        /AA <<%
          /U <<%
            /S/JavaScript/JS(%
              var fd=event.target;%
              var mX=this.mouseX;var mY=this.mouseY;%
              var drag=function(){%
                var nX=this.mouseX;var nY=this.mouseY;%
                var dX=nX-mX;var dY=nY-mY;%
                var fdr=fd.rect;%
                fdr[0]+=dX;fdr[1]+=dY;fdr[2]+=dX;fdr[3]+=dY;%
                fd.rect=fdr;mX=nX;mY=nY;%
              };%
              if(!dragging\thetcnt){%
                dragging\thetcnt=true;Int=app.setInterval("drag()",1);%
              }%
              else{app.clearInterval(Int);dragging\thetcnt=false;}%
              this.dirty=false;%
            )%
          >>%
        >>%
      }%
      \tpAppendToFields{\tpPdfLastAnn}%
    }%
  }}%
  \stepcounter{tcnt}%
}}
\makeatother
\newsavebox\tiptext\newcounter{tcnt}
\newlength{\whatsleft}\newlength{\tipshift}
\newcommand{\measureremainder}[1]{%
  \begin{tikzpicture}[overlay,remember picture]
    \path let \p0 = (0,0), \p1 = (current page.east) in
      [/utils/exec={\pgfmathsetlength#1{\x1-\x0}\global#1=#1}];
  \end{tikzpicture}%
}